\documentclass[reqno]{amsart}
\usepackage{amssymb}
\usepackage{upref}

\newcommand{\lb}{\label}
\newcommand{\bi}{\bibitem}
\newcommand{\ti}{\widetilde}
\newcommand{\bbN}{{\mathbb{N}}}
\newcommand{\bbR}{{\mathbb{R}}}

\DeclareMathOperator{\KdV}{KdV}
\DeclareMathOperator{\mKdV}{mKdV}

\newcommand{\curl}{\operatorname{curl}}

\allowdisplaybreaks
\numberwithin{equation}{section}

\newtheorem{theorem}{Theorem}[section]
\newtheorem{lemma}[theorem]{Lemma}

\theoremstyle{definition}

\theoremstyle{remark}
\newtheorem{remark}[theorem]{Remark}

\begin{document}

\title[Cole-Hopf and Miura Transformations]{The Cole-Hopf 
and Miura transformations  revisited}

\dedicatory{Dedicated with great pleasure to Ludwig Streit on
the occasion  of his 60th birthday} 

\author{Fritz Gesztesy}
\address{Department of Mathematics,
University of Missouri,
Columbia, MO 65211, USA}
\email{fritz@math.missouri.edu}
\urladdr{http://www.math.missouri.edu/people/fgesztesy.html}
\author{Helge Holden}
\address{Department of Mathematical Sciences, Norwegian 
University of 
Science and Technology, N--7034 Trondheim, Norway}
\email{holden@math.ntnu.no}
\urladdr{http://www.math.ntnu.no/\~{}holden/} 

\thanks{Supported in part by the Research Council of 
Norway under grant
107510/410  and the
University of Missouri Research Board grant RB-97-086.}

\begin{abstract}
An elementary yet remarkable similarity between the Cole-Hopf
transformation relating the Burgers and heat equation and
Miura's transformation connecting the KdV and mKdV equations 
is studied in detail.

\end{abstract}

\maketitle

\section{Introduction} \lb{intro}

Our aim in this note is to display the close similarity 
between the 
well-known Cole--Hopf transformation relating the Burgers
 and the heat equation, and the celebrated Miura
transform connecting  the Korteweg--de 
Vries (KdV) and the modified KdV (mKdV) 
equation.  In doing so we will
introduce an additional twist in the Cole--Hopf 
transformation (cf.
\eqref{transfo0}, \eqref{transfo1}), which to the 
best of our 
knowledge, appears to be new. Moreover, we will reveal the 
history of 
this transformation and uncover several instances 
of its 
rediscovery (including those by Cole and Hopf).

We start with a brief introductory account on the KdV and 
mKdV
equations. The KdV equation \cite{KortewegdeVries:1895} 
was
derived as an equation modeling the behavior of 
shallow water waves moving in one
direction  by  Korteweg and his 
student de Vries in 1895\footnote{But the equation 
had been derived
earlier by Boussinesq \cite{Boussinesq:1871} in 
1871, see Heyerhoff
\cite{Heyerhoff:1997} and Pego \cite{Pego:1998}.}.  
The
landmark discovery of the inverse scattering method 
by Gardner, Green,
Kruskal, and Miura in 1967 
\cite{GardnerGreenKruskalMiura:1967} (cf. also
\cite{GGKM74}) brought the KdV equation to the 
forefront of mathematical
physics, and started the phenomenal development 
involving multiple
disciplines of science as well as several branches 
of mathematics.

The KdV equation (in a setting convenient for our purpose) 
reads
\begin{equation}
\KdV(V)=V_t-6 V V_x+V_{xxx}=0, \lb{KdV}
\end{equation}
while its modified counterpart, the mKdV equation, equals
\begin{equation}
\mKdV(\phi)=\phi_t-6 \phi^2 \phi_x+\phi_{xxx}=0. 
\lb{mKdV}
\end{equation}
Miura's fundamental discovery \cite{Miura:1968} 
was the realization that if 
$\phi$ satisfies the
mKdV equation \eqref{mKdV}, then 
\begin{equation}
V_\pm(x,t)=\phi (x,t)^2 \pm\phi_x(x,t), \quad 
(x,t)\in\bbR^2  \lb{Vpm}
\end{equation}
both satisfy the KdV equation. The transformation 
\eqref{Vpm} has since been
called the Miura transformation.   Furthermore, 
explicit calculations by Miura showed
the validity of the identity
\begin{equation}
\KdV(V_\pm)=(2\phi\pm\partial_x)\mKdV(\phi). \lb{miura}
\end{equation}
The Miura transformation \eqref{Vpm} was quite prominently  
used in the
construction of an infinite series of conservation 
laws for the KdV
equation, see 
\cite{MiuraGardnerKruskal:1968}, \cite{DJ89}, Sect.~5.1.
Miura's identity \eqref{miura} then demonstrates how to
transfer  solutions of the mKdV
equation to solutions of the KdV equation, but due 
to the nontrivial
kernel of $(2\phi\pm\partial_x),$ it is not immediately 
clear how to reverse the procedure and to transfer
solutions of the KdV equation to solutions of the mKdV 
equation. It was shown in 
\cite{GesztesySimon:1990} (see also \cite{Ge89}, \cite{Ge91},
\cite{GU92}, \cite{GSS91}) how to revert the process.

Following a similar treatment of the (modified) 
Kadomtsev-Petviashvili
equation in \cite{GU95}, we use here a method that 
considerably simplies the
proofs in
\cite{Ge89}, \cite{GesztesySimon:1990},
\cite{GU92}, \cite{GSS91}. Introduce the first-order 
differential expression
\begin{equation}
\ti P(V)=2V\partial_x-V_x. \lb{P}
\end{equation}
Then one derives 
\begin{equation}
\mKdV(\phi)=\partial_x\left(\frac{1}{\psi}
\big(\psi_t-\widetilde P(V_\pm) \psi
\big)\right), \lb{mmkdv}
\end{equation}
where
\begin{equation}
\phi=\psi_x/\psi, \quad \psi>0,  \quad  V_\pm=
\phi^2\pm\phi_x. \lb{1.6}
\end{equation}
Next, let 
$V=V(x,t)$ be a solution of the KdV equation, 
$\KdV(V)=0$, and 
$\psi>0$ be a function satisfying
\begin{equation}
\psi_t=\ti P(V) \psi, \quad -\psi_{xx}+V \psi=0. \lb{psi}
\end{equation}
Then one immediately deduces that $\phi$ solves the 
mKdV equation, $\mKdV(\phi)=0$, and hence  the
Miura transformation has been  ``inverted''.

The KdV equation \eqref{KdV} and the mKdV equation 
\eqref{mKdV} are 
just the first
(nonlinear) evolution equations in a countably 
infinite hierarchy of 
such equations (the
(m)KdV hierarchy).  The considerations 
\eqref{Vpm}--\eqref{psi} extend 
to the entire
hierarchy of these equations, replacing the first-order 
differential 
expression $\widetilde P(V)=\widetilde P_1(V)$
by an appropriate first-order differential expression 
$\widetilde P_n(V)$ for 
$n\in\bbN$ (cf., e.g.,
\cite{Ge89}, \cite{Ge91}, \cite{GU92}, \cite{GSS91}). 
More precisely, denoting the $n$th KdV equation in the 
KdV hierarchy by 
\begin{equation}
\KdV_n(V)=0, \lb{Kn}
\end{equation}
Lax \cite{La68} constructed differential expressions
$P_{2n+1}(V)$ of order $2n+1$ with coefficients differential 
polynomials of $V$ such that
\begin{equation}
\frac{d}{dt}L - [P_{2n+1}(V),L]=\KdV_n(V), \quad n\in\bbN. 
\lb{KdV_n}
\end{equation}
Here $L$ denotes the Schr\"odinger differential expression 
\begin{equation}
L=-\partial^2_x + V. \lb{L}
\end{equation}
The KdV functional in \eqref{KdV} then corresponds to $n=1$
and one obtains
\begin{equation}
P_3(V)= -4\partial^3_x + 6V\partial_x + 3V_x \lb{P3}
\end{equation}
in this case. Restriction of $P_{2n+1}(V)$ to the (algebraic)  
nullspace of $L$ then yields the first-order differential 
expression 
\begin{equation}
\ti P_n(V) = P_{2n+1}(V)\big|_{\ker(L)}, \quad n\in\bbN. 
\lb{tPn}
\end{equation}

Next we  turn to the the Cole--Hopf 
transformation and its history. The classical Cole--Hopf
transformation 
\cite{Cole:1951}, 
\cite{Hopf:1950}, covered in 
most textbooks on partial differential equations, states that 
\begin{equation}
V(x,t)=-2\frac{\psi_{x}(x,t)}{\psi(x,t)},	\quad
(x,t)\in\bbR\times (0,\infty), 
\end{equation}	
where $\psi >0$ is a
solution  of the heat equation
\begin{equation}
\psi_{t}=\psi_{xx},	
\end{equation}
satisfies the (viscous) Burgers equation
\begin{equation}
V_{t}+VV_x=V_{xx}.	
\end{equation}
However, already in 1906, Forsyth, in his multi-volume 
treatise on 
differential equations (\cite{Forsyth:1906}, p.\ 100), 
discussed the equation
(in his notation)
\begin{equation}
\frac{\partial}{\partial x}
\left\{\frac{1}{\beta}\left(\gamma-
\frac{\partial\alpha}{\partial 
x}-\alpha^2  \right)  \right\}-
2\frac{\partial\alpha}{\partial 
y}=0,	\lb{forsyth1}
\end{equation}
where $\alpha=\alpha(x,y)$.  Hence there exists 
a function $\theta$ 
such that
\begin{equation}
	\alpha=\frac{\partial\theta}{\partial x}, \quad
	\gamma-\frac{\partial\alpha}{\partial x}-\alpha^2
	=2\beta\frac{\partial\theta}{\partial y}. 
\end{equation}
Assuming the function $z$ satisfies
\begin{equation}
z_{xx}+2\alpha z_{x}+2\beta z_{y} + \gamma =0,	
\end{equation}
an easy calculation shows that
\begin{equation}
\frac{\partial^2}{\partial x^2}(z e^\theta)
+2\beta\frac{\partial}{\partial y}(z e^\theta)=0. 
\lb{forsyth2}	
\end{equation}
Introducing new variables $t=-y$ and $u(x,t)=
-2 \alpha(x,y)$ 
as well as fixing $\beta=1/2$, $\gamma=0$, and $z=1$,
 one concludes that  \eqref{forsyth1} indeed 
reduces to the viscous Burgers equation 
\begin{equation}
u_t+uu_x=u_{xx},
\end{equation}
while \eqref{forsyth2} equals
\begin{equation}
(e^\theta)_t=(e^\theta)_{xx},
\end{equation}
with solutions related by
\begin{equation}
u=-2\theta_x.
\end{equation}
However,  Forsyth did not study the ramifications of 
this 
transformation, and no applications are discussed.

Shortly thereafter, in 1915, Bateman \cite{Bateman:1915} 
introduced
the model equation
\begin{equation}
u_{t}+u u_{x}=\nu u_{xx}. \lb{bateman}
\end{equation}
He was interested in the vanishing viscosity limit, 
that is, when 
$\nu\to 0$.  By studying solutions of the form 
$u=F(x+Ut)$, he
concluded that ``the question of the limiting form 
of the motion of a 
viscous fluid when the viscosity tends to zero requires 
very careful 
investigation''.

Only in 1940 did Burgers (\cite{Burgers:1940}, p.\ 8)
introduce\footnote{Frequently Burgers equations is 
quoted from his
1948 paper \cite{Burgers:1948}, but he had already 
introduced it in
1940.} what has later 
been called the (viscous) Burgers equation, as a 
simple model of
turbulence, and did some preliminary investigation 
on properties of
the solution.

Taking advantage of the later rediscovered Forsyth  
transformation by Cole and Hopf,
 Burgers continued the investigations of what he 
called the nonlinear
 diffusion equation, focusing mainly on statistical 
aspects of the
 equation. The results of these investigations
were collected in his book \cite{Burgers:1974}.

In 1948, Florin \cite{Florin:1948}, in the context 
of applications to
watersaturated flow, rediscovered Forsyth's 
transformation, which would become well-known under 
the name Cole-Hopf transformation only some 44 years later.

Although the Cole--Hopf transformation had already been 
published in
1906, it was only with the seminal papers by Hopf 
\cite{Hopf:1950}\footnote{With a misprint in the title,
writing $u_t+u u_x=\mu_{xx}$ rather than 
$u_t+u u_x=\mu u_{xx}$.} 
in 1950\footnote{Hopf \cite{Hopf:1950}
states in a footnote (p.\ 202) that he had the 
``Cole--Hopf
transformation'' already in 1946, but ``it was not 
until 1949 that I
became sufficiently acquainted with the recent 
development of fluid
dynamics to be convinced that a theory based on 
\eqref{bateman} could
serve as an instructive introduction into some of 
the mathematical
problems involved''.}
and by Cole \cite{Cole:1951} in 1951  that the full
impact of the simple transformation was seen.  In 
particular the
careful study by Hopf concerning the vanishing 
viscosity limit
represented a landmark in the emerging theory of 
conservation laws.
Although the Cole--Hopf transformation is restricted 
to the Burgers
equation, the insight and the motivation from this 
analysis has been
of fundamental importance in the theory of 
conservation laws.
Furthermore, Cole states the generalization of the 
Cole--Hopf
transformation to a particular multi-dimensional system. 
More precisely,
if $\psi=\psi(x,t)$, $(x,t)\in\bbR^n\times (0,\infty)$,
satisfies  the $n$-dimensional 
heat equation
\begin{equation}
\psi_t=\nu\Delta\psi, \quad \nu>0, \lb{cole1}
\end{equation}
and one defines
\begin{equation}
V=-2\nu\nabla\ln(\psi), \lb{cole2}
\end{equation}
then $V$ satisfies
\begin{equation}
V_t+(V\cdot\nabla) V=\nu \Delta V, \lb{cole3}
\end{equation}
and the vector-valued function
$V=V(x,t)\in\bbR^n$ has as many components (i.e., $n$) 
as the dimension of the
underlying  space.  Observe, in
particular, that $V$ is irrotational (i.e., $V=\nabla W$
for some $W,$ or equivalently, $\curl V=0,$).
The multi-dimensional extension was rediscovered 
by Kuznetsov and
Rozhdestvenskii \cite{KuznetsovRozhdestvenskii:1961} 
in 1961.

In this note we  show the following relations,
\begin{align}
V_t+VV_x-\nu V_{xx}& =
2\nu\left(- \frac{1}{\psi}\partial_x
+\frac{\psi_x}{\psi^2} \right)\left(\psi_t-
\nu\psi_{xx}\right)\lb{transfo0}\\
&=-2\nu\partial_x\left(\frac{1}{\psi}
(\psi_t-\nu\psi_{xx}) \right), \lb{transfo1}
\end{align}
whenever $V=-2\nu \psi_x/\psi$ for a positive function 
$\psi.$ This 
clearly displays the nature of the Cole--Hopf transformation 
and 
closely resembles Miura's identity
\eqref{miura} and the relation \eqref{mmkdv}. Even though
identities \eqref{transfo0} and \eqref{transfo1} are 
elementary observations, much to our surprise, they appear 
to have escaped
notice in the extensive literature on the Cole-Hopf 
transformation thus
far. While both the KdV and mKdV equations are nonlinear
partial differential equations, the case of the Burgers and
heat equations just considered is a bit different since it 
relates a nonlinear and a linear partial differential 
equation (see also \cite{BK89}, Sect.~6.4).

One can also extend the Cole--Hopf transformation to 
the case of a
potential term $F$ in the heat equation, see, for instance,
\cite{HoldenOksendalUboeZhang:1996}.
Here  the  relation \eqref{transfo1} reads as follows,
\begin{equation}
V_t+VV_x-\nu V_{xx}+2\nu F_x 
=-2\nu\partial_x\left(\frac{1}{\psi}
(\psi_t-\nu\psi_{xx}-F\psi) \right), \lb{transfo2}
\end{equation}
whenever $V=-2\nu \partial_x \ln(\psi)$ for a positive 
function $\psi$. The case of Burgers' equation externally 
driven by a random potential term recently generated 
particular
interest, see, for instance, \cite{BS98}, \cite{BDPS98},
\cite{GKS98}, \cite{HoldenOksendalUboeZhang:1996}, 
\cite{HLOUZ94}, \cite{HLOUZ95} and the 
references therein. We
also  mention a very interesting application of the 
Cole-Hopf
transformation to the pair of the telegraph and a 
nonlinear 
Boltzmann equation in \cite{HS90}, generalizing the 
pair of the
heat and  Burgers equation considered in this note.

Equation \eqref{transfo2} extends to the  
multi-dimensional
case corresponding to \eqref{cole3} and one obtains
\begin{equation}
V_t+\alpha (V\cdot\nabla) V-\nu \Delta V+
\frac{2\nu}{\alpha}\nabla F 
=-\frac{2\nu}{\alpha}\nabla\left(\frac{1}{\psi}
\big(\psi_t-\nu\Delta\psi-F\psi\big)\right), 
\end{equation}
whenever
$\alpha\in\bbR\backslash\{0\}$
and $V=-(2\nu/\alpha)\nabla \ln(\psi)$ for a  positive
function $\psi$.

Obviously there is a close similarity between the
heat and the Burgers
 equation expressed by  \eqref{transfo0}, and Miura's
identity 
\eqref{miura} relating the mKdV and the KdV equation.  
The principal idea underlying these
considerations being that one (hierarchy of) evolutions 
equation(s) can be
represented as a linear differential expression acting on 
another (hierarchy of)
evolution equation(s).  As long as the null space of this 
linear differential
expression can be analyzed in detail, it becomes possible 
to transfer solutions,
in fact, entire classes of solutions (e.g., rational, 
soliton, 
algebro-geometric
solutions, etc.) between these evolution equations. 
In concrete
applications, however, it turns out to be simpler to 
rewrite a relationship
between two evolution equations, such as
\eqref{miura} and \eqref{transfo0}, in a form analogous to
\eqref{mmkdv} and \eqref{transfo1}, rather than 
analyzing the
nullspaces of $(2\phi\pm \partial_x)$ and $(-\partial_x 
+ (\psi_x/\psi))$ in
detail.  These  strategies relating 
(hierarchies of) evolution equations and their modified 
analogs is not at all restricted to the Burgers
and heat equations and the KdV and mKdV hierarchies, 
respectively, but applies to a large
number of evolution equations  including the 
Boussinesq \cite{GRW93}, and more
generally, the Gelfand--Dickey hierarchy and its modified 
counterpart, the
Drinfeld--Sokolov hierarchy
\cite{GRUW94}, the Toda and Kac--van Moerbeke hierarchies 
\cite{BGHT98}, \cite{GHSZ93}, 
the Kadomtsev--Petviashvili and modified 
Kadomtsev--Petviashvili 
hierarchies \cite{GU95}, \cite{GHSS91}, etc.

For simplicity we restrict ourselves to classical solutions
throughout this note. The case of distributional solutions 
of Burgers equation is considered, for instance, 
in \cite{Di96}.

Throughout this note we abbreviate by
$C^{p,q}(\Omega\times\Lambda),$ 
$\Omega\subset\bbR^n, \Lambda\subset\bbR$ open, $n\in\bbN,$
$p,q\in\bbN_0,$ the linear space  of continuous functions
$f(x,t)$ with continuous  partial derivatives with
respect to $x=(x_1,\dots,x_n)$ up to order $p$ and 
$q$ partial 
derivatives with respect to $t.$
$C^{p,q}(\Omega\times\Lambda ;\bbR^n)$ is then defined
analogously for $f(x,t)\in\bbR^n.$


\section{The Miura transformation} \lb{KdV-section}


We turn to the precise formulation of the relations 
between the KdV and the mKdV equation and omit details of a 
purely calculational nature. 

\begin{lemma}
\lb{l3.1}
Let $\psi=\psi(x,t) >0$ be a positive function such that
$\psi\in
C^{4,0}(\bbR\times\bbR),$ $\psi_t \in
C^{1,0}(\bbR\times\bbR).$   Define
$\phi=\psi_x/\psi$. Then $\phi\in  
C^{3,1}(\bbR\times\bbR)$ and
\begin{equation}
\mKdV(\phi)=\partial_x\left(\frac{1}{\psi}
\big(\psi_t-\ti P(V_\pm) \psi
\big)\right), \lb{karl}
\end{equation}
where 
\begin{equation}
\ti P(V) =2V\partial_x-V_{x} \lb{B}
\end{equation}
and
\begin{equation}
V_\pm=\phi^2\pm\phi_x.
\end{equation}
\end{lemma}
\begin{proof}
A straightforward calculation.
\end{proof}

The application to the KdV equation then reads as follows.

\begin{theorem} \lb{t3.2}
Let $V=V(x,t)$ be a solution of the KdV equation, 
$\KdV(V)=0,$ with $V\in
C^{3,1}(\bbR\times \bbR)$, and let
$\psi >0$ be a positive function satisfying 
$\psi\in C^{2,0}(\bbR\times\bbR),$ $\psi_t \in
C^{1,0}(\bbR\times\bbR)$ and
\begin{equation}
\psi_t=\ti P(V) \psi, \quad -\psi_{xx}+V \psi=0, 
\lb{syst}
\end{equation}
with $\ti P(V)$ given by \eqref{B}.  Define 
$\phi=\psi_x/\psi$ and $\widehat V =\phi^2-\phi_x$.  
Then $V=\phi^2+\phi_x$ and $\phi$ satisfies 
$\phi\in C^{4,1}(\bbR\times\bbR)$ and the mKdV equation,  
\begin{equation}
\mKdV(\phi)=0.
\end{equation} 
Moreover,  $\widehat V$ satisfies $\widehat V\in
C^{3,1}(\bbR\times
\bbR)$ and the KdV equation,
\begin{equation}
\KdV(\widehat V) =0. 
\end{equation}
\end{theorem}

\begin{proof}
A computation based on Lemma~\ref{l3.1}.
\end{proof}

Originally, Theorem~\ref{t3.2} was proved in
\cite{GesztesySimon:1990} (see also \cite{Ge89}, \cite{Ge91},
\cite{GS95}, \cite{GU92}, \cite{GSS91}) using supersymmetric 
methods. The above arguments, following
\cite{GU95} in the  context of the (modified)
Kadomtsev-Petviashvili equation,  result in considerably
shorter calculations. The ``if part'' in Theorem~\ref{t3.2} 
also follows from prolongation methods developed in 
\cite{WE75}. A different approach to Theorem~\ref{t3.2}, 
assuming rapidly decreasing solutions of the KdV equation, 
can be found in Sect.~38 of \cite{BDT88}.
 
\begin{remark} \lb{r2.3}
The chain of transformations
\begin{equation}
V\to \phi \to -\phi \to \widehat V \lb{BT}
\end{equation}
reveals a B\"acklund transformation between the KdV and mKdV 
equations ($V\to\phi$) and two auto-B\"acklund 
transformations 
for the KdV ($V\to\widehat V$) and mKdV equations ($\phi\to
-\phi$), respectively.
\end{remark}
\begin{remark} \lb{r2.4}
For simplicity we assumed $\psi(x,t)>0$ for all 
$(x,t)\in\bbR^2$ in Theorem~\ref{t3.2}. However, as proven by
Lax \cite{La74}, one can show that $\psi(x,t_0)>0$ for some 
$t_0\in\bbR$ and all $x\in\bbR$ actually implies 
$\psi(x,t)>0$ for all $(x,t)\in\bbR^2$ (see also 
\cite{GesztesySimon:1990}). Moreover, in case $V(x,t_0)$ is 
real-valued, we note that 
$L(t_0)\psi(x,t_0)=0$ has a positive solution $\psi(x,t_0)>0$ 
if and only if the Schr\"odinger differential expression 
$L(t_0)=-\partial^2_x + V(x,t_0)$ is nonoscillatory at 
$\pm\infty$ (cf.~\cite{GZ91}). While the system of equations 
\eqref{syst} always has a solution $\psi(x,t),$ 
(cf.~Lemma~3 in 
\cite{GesztesySimon:1990}), it is the additional requirement 
$\psi(x,t)>0$ for all $(x,t)\in\bbR^2$ which renders $\phi$ 
(and hence $\widehat V$) nonsingular. Without the condition 
$\psi >0,$ Theorem~\ref{t3.2} describes (auto)B\"acklund 
transformations for the KdV and mKdV equations with
characteristic  singularities (cf.~\cite{GS95}). 
\end{remark}

\section{The Cole--Hopf transformation} \lb{Burgers-section}

Finally we return to relations \eqref{transfo0},
\eqref{transfo1}, and \eqref{transfo2}. Since they are 
all proved by explicit calculations we may omit these details 
and focus on a precise formulation of the results
instead. 

\begin{lemma} \lb{gh1}
Let $\psi=\psi(x,t) >0$ be a 
positive function with $\psi\in C^{3,0}(\bbR\times
(0,\infty)),$ $\psi_t\in
C^{1,0}(\bbR\times  (0,\infty)).$ Define
$V=-2\nu \psi_x/\psi$ with $\nu >0$.  Then $V\in
C^{2,1}(\bbR\times (0,\infty))$ and
\begin{align}
V_t+VV_x-\nu V_{xx}& =
2\nu\left(- \frac{1}{\psi}\partial_x
+\frac{\psi_x}{\psi^2} 
\right)\left(\psi_t-\nu\psi_{xx}\right)
\lb{ch1}\\
&=-2\nu\partial_x\left(\frac{1}{\psi}
(\psi_t-\nu\psi_{xx}) \right). \lb{ch2}
\end{align}
\end{lemma}

The extension to the case with a potential term $F$ in the 
heat equation 
reads as follows.

\begin{lemma} \lb{gh1a}
Let $F\in C^{1,0}(\bbR\times(0,\infty))$ and assume 
$\psi=\psi(x,t) >0$ to be a 
positive function such that $\psi\in C^{3,0}(\bbR\times
(0,\infty)),$ $\psi_t \in 
C^{1,0}(\bbR\times (0,\infty)).$  Define
$V=-2\nu \psi_x/\psi$ with $\nu >0$.  Then $V\in
C^{2,1}(\bbR\times  (0,\infty))$ and 
\begin{equation}
V_t+VV_x-\nu V_{xx}+2\nu F_x 
=-2\nu\partial_x\left(\frac{1}{\psi}
(\psi_t-\nu\psi_{xx}-F\psi) \right). \lb{ch22}
\end{equation}
\end{lemma}

We can exploit these relations as follows.

\begin{theorem} \lb{gh3}
Let $F\in C^{1,0}(\bbR\times (0,\infty))$ and $\nu>0.$\\
(i) Suppose $V$ satisfies $V\in C^{2,1}(\bbR\times 
(0,\infty))$, and
\begin{equation}
V_t+VV_x-\nu V_{xx}+2\nu F_x =0
\lb{ikkelin}
\end{equation}
for some $\nu >0.$ Define
\begin{equation}
\psi(x,t)=\exp\left(-\frac1{2\nu} \int^x dy\, V(y,t)\right).
\end{equation}
Then $\psi$ satisfies $0<\psi\in C^{3,1}(\bbR\times 
(0,\infty))$ and
\begin{equation}
\frac1{\psi}\left(\psi_t-\nu \psi_{xx}- F \psi \right)=C(t)
\end{equation}
for some $x$-independent $C\in C(\bbR)$. \\
(ii) Let $\psi >0$ be a positive function satisfying 
$\psi\in
C^{3,0}(\bbR\times (0,\infty)),$ $\psi_t \in
C^{1,0}(\bbR\times(0,\infty))$ and suppose
\begin{equation}
\psi_t=\nu \psi_{xx}+F \psi 
\end{equation}
for some $\nu >0.$ Define
\begin{equation}
V=-2\nu \frac{\psi_x}{\psi}.
\end{equation}
Then $V\in C^{2,1}(\bbR \times (0,\infty))$ satisfies
\eqref{ikkelin}.
\end{theorem}

\begin{remark} \lb{r3.4} 
One can ``scale away'' $C(t)$ in 
Theorem~\ref{gh3}\,(i) by 
introducing a new function $\tilde \psi$.
In fact, the function  $\tilde \psi(x,t)=
\psi(x,t)\exp(-\int^t_0 ds\, C(s))$ satisfies
\begin{equation}
\tilde\psi=\nu \tilde\psi_{xx}+F \tilde\psi . 
\end{equation}
\end{remark}
\begin{remark} \lb{r3.5}
Using the standard representation of solutions of the
heat equation initial value problem,
\begin{equation}
\psi_t=\nu\psi_{xx}, \quad \psi(x,0)=\psi_0(x), 
\end{equation}
assuming
\begin{equation}
\psi_0\in C(\bbR), \quad \psi_0(x)\leq
C_1\exp(C_2|x|^{1+\gamma}) \text{ for } |x|\geq R
\end{equation}
for some $R>0,$ $C_j\geq 0,$ $j=1,2,$ and $0\leq\gamma < 1,$
given by (cf.~\cite{Ca84}, Ch.~3; \cite{Di95}, Ch.~V)
\begin{equation}
\psi(x,t)=\frac{1}{2(\pi\nu t)^{1/2}}\int_\bbR dy \,
\exp(-(x-y)^2/4\nu t) \psi_0(y)>0,
\end{equation}
the corresponding initial value problem for the Burgers 
equation 
\begin{equation}
V_t+VV_x -\nu V_{xx}=0, \quad 
V(x,0)=V_0(x),
\end{equation}
reads
\begin{equation}
V(x,t)=\frac{\int_\bbR dy \, (x-y)t^{-1}\exp\big(-(2\nu)^{-1}
\int_0^y d\eta V_0(\eta) - (x-y)^2(4\nu t)^{-1} \big)}
{\int_\bbR dy \, \exp\big(-(2\nu)^{-1}
\int_0^y d\eta V_0(\eta) - (x-y)^2(4\nu t)^{-1} \big)},
\end{equation}
assuming $V_0\in C(\bbR)$ and 
\begin{equation}
V_0(x)\geq 0 \text{ or } \bigg|\int_0^x dy \, V_0(y)\bigg|
\underset{|x|\to\infty}=
O(|x|^{1+\gamma}) \text{ for some } \gamma <1.
\end{equation}
\end{remark}
\vspace*{3mm}

Without going into further details we mention the 
existence of
a hierarchy of Burgers equations which are related to the
linear partial differential equations
\begin{equation}
\psi_t=\partial^{n+1}_x\psi, \quad n\in\bbN
\end{equation}
by the Cole-Hopf transformation $V=-2\psi_x/\psi$ (see 
\cite{Bl86}).\\

The multi-dimensional
extension of Lemma~\ref{gh1a}  reads as follows.

\begin{lemma} \lb{gh2}
Let $F\in C^{1,0}(\bbR^n\times(0,\infty))$ and assume 
$\psi=\psi(x,t) >0$ to be a positive function such that 
$\psi\in
C^{3,0}(\bbR^n\times (0,\infty)),$ $\psi_t\in
C^{1,0}(\bbR^n\times (0,\infty)).$ Define
$V=-(2\nu/\alpha)\nabla \ln(\psi)$ with 
$\alpha\in\bbR\backslash\{0\},$ $\nu >0.$ Then 
$V\in C^{2,1}(\bbR^n\times(0,\infty);\bbR^n)$ and
\begin{equation}
V_t+\alpha (V\cdot\nabla) V-\nu \Delta V+
\frac{2\nu}{\alpha}\nabla F 
=-\frac{2\nu}{\alpha}\nabla\left(\frac{1}{\psi}
\big(\psi_t-\nu\Delta\psi-F\psi\big)\right). 
\end{equation}
\end{lemma}

Our final result shows how to transfer solutions between 
the multi-dimensional Burgers equation and the heat 
equation.

\begin{theorem} \lb{gh5}
Let $F\in C^{1,0}(\bbR^n\times(0,\infty)),$ 
$\alpha\in\bbR\backslash\{0\},$ and $\nu >0.$ \\
(i) Assume  that $V\in C^{2,1}(\bbR^n\times 
(0,\infty);\bbR^n)$ 
satisfies 
\begin{equation}
V=\nabla \Phi
\end{equation}
for some potential $\Phi\in C^{3,1}(\bbR^n\times 
(0,\infty))$ 
and
\begin{equation}
V_t+\alpha (V\cdot  \nabla) V-
\nu \Delta V+\frac{2\nu}{\alpha}\nabla F =0. \lb{lign}
\end{equation}
Define 
\begin{equation}
\psi=\exp\left(-\frac{\alpha}{2\nu}\Phi \right).
\end{equation}
Then $\psi\in C^{3,1} (\bbR^n\times (0,\infty))$ and 
\begin{equation}
\frac{1}{\psi}
\big(\psi_t-\nu\Delta\psi-F\psi\big)=C(t),
\end{equation}
for some $x$-independent $C\in C((0,\infty)).$ \\
(ii) Let $\psi >0$ be a positive function satisfying  
$\psi\in
C^{3,0}(\bbR^n\times (0,\infty)),$ $\psi_t\in
C^{1,0}(\bbR^n\times (0,\infty))$ and suppose
\begin{equation}
\psi_t=\nu\Delta\psi+ F\psi.
\end{equation}
Define 
\begin{equation}
V=-\frac{2\nu}{\alpha}\nabla\ln(\psi).
\end{equation}
Then $V\in C^{2,1}(\bbR^n\times(0,\infty);\bbR^n)$ 
satisfies \eqref{lign}.
\end{theorem}

\bigskip

{\bf Acknowledgments.}   We thank Mehmet \"Unal 
and Karl Unterkofler
for discussions on the Burgers and (m)KdV equations, 
respectively.


\end{document}